# Machine Learning-Based Battery State-of-health Prediction for Unmanned Aerial Vehicles Predictive Maintenance

Jiarui Xie
Department of Mechanical Engineering
McGill University
Montreal, Canada
jiarui.xie@mail.mcgill.ca

Lingchen Kong
Department of Mechanical Engineering
McGill University
Montreal, Canada
lingchen.kong@mail.mcgill.ca

Mohamed Rami Latreche
Department of Information Systems and Quantitative Management Methods
Université de Sherbrooke
Sherbrooke, Canada
mohamed.rami.latreche@usherbrooke.ca

Sean Smith
VOZWIN Inc.
Pointe-Claire, Canada
smith.sean@vozwin.com

Elaine Mosconi
Department of Information Systems and Quantitative Management Methods
Université de Sherbrooke
Sherbrooke, Canada
elaine.mosconi@usherbrooke.ca

Yaoyao Fiona Zhao
Department of Mechanical Engineering
McGill University
Montreal, Canada
yaoyao.zhao@mcgill.ca

***Abstract*— Battery state-of-health (SoH) prediction aims to estimate the remaining capacity by modeling battery degradation through its life cycle. Machine learning (ML)-based SoH models can accurately predict the battery remaining capacity based on voltage, current, and temperature. Battery SoH prediction for unmanned aerial vehicles (UAVs) is a crucial yet overlooked domain with data scarcity and high variability. Accurate battery SoH information contributes to efficient predictive maintenance, enhancing UAV profitability and flight safety. However, UAVs are compatible with a variety of batteries and the available data for each type of battery are scarce. Furthermore, the available input features from UAV batteries are limited to the built-in sensors because of the lightweight requirements. This research aims to develop an ML pipeline for UAV battery SoH prediction while mitigating data scarcity using knowledge transfer. 342 and 289 flight experiments have been conducted to collect operational data from lithium polymer batteries of 2200 mAh and 1100 mAh, respectively. Voltage, current, and throttle are selected as the input features of the ML model according to the existing literature and sensor availability. The remaining capacity is measured at every 10th experiment to label the dataset. To address data scarcity, the time-series data acquired from the experiments are transformed into images to utilize the image feature extraction capability of a pretrained ResNet-50. Finally, accurate SoH prediction models were obtained using transfer learning between two battery types.**

***Keywords*— *Machine learning, predictive maintenance, battery, state-of-health, unmanned aerial vehicle***

## I. INTRODUCTION AND BACKGROUND

Lithium batteries, which feature high energy density, long life span, and cost-efficiency, have become a cornerstone in the field of energy storage technology and electromobility, particularly revolutionizing the electric vehicles (EVs) and unmanned aerial vehicles (UAVs) [1, 2]. However, the degradation of lithium batteries along the charging and discharging cycles increases the risk of vehicle accidents [3]. Predictive Maintenance (PdM) is widely implemented for UAVs and EVs to enhance their reliability and extend the overall lifespan. Different from traditional maintenance technologies, PdM can monitor the system state of health (SoH) and predict potential faults before they occur. PdM continuously monitors and processes the machine operational data to estimate the health indicators (HIs), SoH, and remaining useful life (RUL), which are the most critical parameters of lithium batteries [4-6].

Machine Learning (ML) is a promising computational intelligence method with unparalleled data analysis capabilities. ML models can discover and extract the underlying degradation patterns from the historical data during operations to support SoH prediction. Hence, many research works have been training and deploying ML models in the PdM of various systems such as gas turbines and 3D printers.

Battery SoH can be reflected by the remaining capacity and internal resistance, which are difficult to measure in real applications [2]. To acquire the run-to-failure data for ML-based SoH prediction, life cycle tests are designed and implemented to iteratively charge and discharge batteries until they reach end-of-life conditions. Throughout the experiments, battery parameters such as voltage, current, and temperature are measured and recorded as the ML input data to conduct SoH estimation.

Battery SoH prediction for UAVs has not been extensively researched compared with EVs. UAV battery data collection is complicated because UAVs are compatible with various types

Mathematics of Information Technology and Complex Systems (MITACS) Accelerate program [grant number IT13369] and Réseau SDG Innovation Network - Collaborative R&D projects in digital, intelligent and sustainable transformation.ere. If none, delete this text box.

of batteries, each with unique characteristics like capacity and working voltage. Different pilot flying habits and diverse environments add to the complexity, making it challenging to create a single ML model suitable for all scenarios. Therefore, the SoH prediction model demands a large capacity and a diverse dataset to account for high variability. Furthermore, the lightweight requirement of UAVs is a constraint on the number of sensors, which limits the available input variables.

To address the abovementioned challenges, this work develops an ML pipeline for UAV battery SOH prediction. Battery life cycle tests were carried out for two types of lithium polymer (LiPo) batteries with different nominal capacities. Knowledge transfer between the two types of batteries was conducted to address data scarcity issues. The enhanced prediction performance demonstrated the effectiveness of the proposed strategy and ML model with limited data size.

ML-based SoH prediction essentially models the relationship between the input variables and the health status. Compared with physics-based and rule-based methods, ML can efficiently extract representative patterns from the input data of various formats and derive degradation signatures without using physics knowledge. For example, once the voltage and current curves during operation are converted to images, convolutional neural networks (CNNs) can automatically extract abstract features to estimate battery health [7-9]. Numerous ML-based time-series forecasting methods have been implemented to model battery degradation. Among them, long short-term memory (LSTM) networks are widely used to accurately estimate battery SOH due to their ability to prevent gradient vanishing and to extract both long-term and short-term information [2, 3, 10, 11]. By combining CNN and LSTM, spatiotemporal features can be derived to predict the battery remaining capacity [12-14].

## II. Methodology

This section discusses the data acquisition, data handling, and ML model training procedures in this research to build a data-driven SoH prediction model for UAV batteries. First, the experiments are designed and implemented to simulate a range of UAV operations with respect to different total weights and flight lengths. Thereafter, the data handling techniques that transform the acquired raw data to facilitate the subsequent image analysis are illustrated. Finally, ML models are trained and compared to select the optimal model and demonstrate the effectiveness of the proposed strategy.

### *A. UAV Battery Experiments*

Similar to other data-driven battery SOH research, this research conducted life cycle tests that continuously charge and discharge the batteries until they reach end-of-life conditions. The end-of-life conditions are when any battery cells in the pack fail or when the remaining capacity decreases to 80% of the initial capacity [3]. To adapt to the practical UAV predictive maintenance setting, batteries are discharged using a UAV, whose configuration is shown in Table I. The experiments are designed to cover a wide range of total weights (150 g to 250 g) and flight lengths (5 minutes to 15 minutes) for a typical lightweight UAV with a 3-cell LiPo battery pack. LiPo batteries are selected because their lightweight and reliable characteristics make them suitable for UAV operations. The flight controller, configured with Pixhawk 4 mini, can collect and record various parameters measured using onboard sensors, including the output voltage and current. The sampling frequencies of the measurements are tuned to 10 Hz. These parameters will be used as the input variables of the ML model to predict the SOH. After each experiment, the battery pack is fully charged using a B6 V2 balance charger, which can measure the capacity charged. The same charging protocol is deployed after every experiment: the battery is first charged using a constant current of 2.2 V, then the charging current gradually decreases with a constant voltage to prevent overcharging. Data collection during the charging cycles of UAV batteries is usually not available and thus was not conducted in the experiments. The experiments aimed to simulate normal UAV flights that perform take-off, steady flight, and landing.

TABLE I. QUADCOPTER CONFIGURATION

| **Component** | **Description** |
|---|---|
| Electronic speed controller (ESC) | Tekko32 F4 4in1 50A ESC |
| Motors | RCTIMER, 4*HP1806-2300kv |
| Flight controller | Holybro, Pixhawk 4 Mini |
| Transmitter | RadioMaster, TX16S Mark II Radio Controller |

In this work, the remaining capacity of the battery is considered the SOH indicator. The experiments are designed to measure the remaining capacity as the output variable of the ML model. However, the remaining capacity can be measured only when the battery is fully discharged, which significantly impairs the battery health. Therefore, the remaining capacity was measured after every 10 successive experiments with a battery tester that discharges the battery to the minimum allowable voltage using the same discharging protocol. The 10 successive experiments between two full discharge cycles employ the same total weight (represented by the same throttle) and the same flight length. It is assumed that the 10 successive experiments have similar degradation effects to the battery SOH. Thus, the remaining capacity of each experiment can be obtained by linearly interpolating between the remaining capacities of the two surrounding full discharge cycles.

LiPo batteries are commonly utilized as the power source of lightweight UAVs but lack existing SOH experimental data. Due to the wide compatibility between UAVs and LiPo batteries, users are likely to have multiple types of batteries with different brands, sizes, and capacities. The experiments conducted in this research involved two types of LiPo batteries with different capacities (shown in Table II). Although more 1100 mAh batteries were tested than 2200 mAh batteries, more experiments were obtained from 2200 mAh batteries. This was because it usually takes more cycles for batteries with larger capacities to reach end-of-life conditions.

*B. Data Handling*

From the experiments, various input variables are measured and recorded during the discharging cycles as time-series data. Relevant literature indicates that the voltage, current, and temperature during the discharging cycles embed crucial battery SOH information [15]. Battery temperature measurements involving extra sensors are not considered due to the lightweight requirements of UAVs. Throttle is added to the input variables as it reflects the drone operations and the associated demand for electricity. Using the time-series data, state-of-the-art time-series forecasting such as transformers are often trained to predict battery SOH. However, the capability of the advanced time-series forecasting methods, which usually employ a significant number of parameters, is significantly compromised due to data scarcity. In this case, advanced ML models are more prone to overfitting, leading to unsatisfactory test accuracy. To address the data scarcity challenge, this paper proposes to convert time-series data to image data, from which pre-trained ML models can be applied to extract salient features (Figure 1(a)). Those extracted features, embedding important information on battery health, are used as part of the input features to train an ML-based SOH prediction model (Figure 1(b)). This way, the number of trainable parameters can be reduced because of the image feature extraction capability of the pre-trained model.

The original time-series data of each variable from each experiment were stored as a 1-dimensional array, where measurements were registered sequentially. Each 1-dimensional array was first normalized to a range of [0, 255], then reshaped to a 2-dimensional matrix with a column size of 112. To represent these 2-dimensional matrices using images, the number in each cell was rounded to the nearest integer. This way, each grayscale image can represent the variation of one input variable during an experiment. Although the data resolution is reduced during the conversion, the new datasets can represent small voltage and current variations of up to 0.05 V and 0.0075 A, respectively. Thereafter, the three grayscale images from one experiment were stacked to construct a 3-channel image. After conversion, the images from different experiments had different sizes due to the varying experiment lengths. Black padding was added to the images to ensure the same image sizes across the dataset. Afterward, the images were resized to 80 pixels by 80 pixels, which is the input data size for ResNet-50

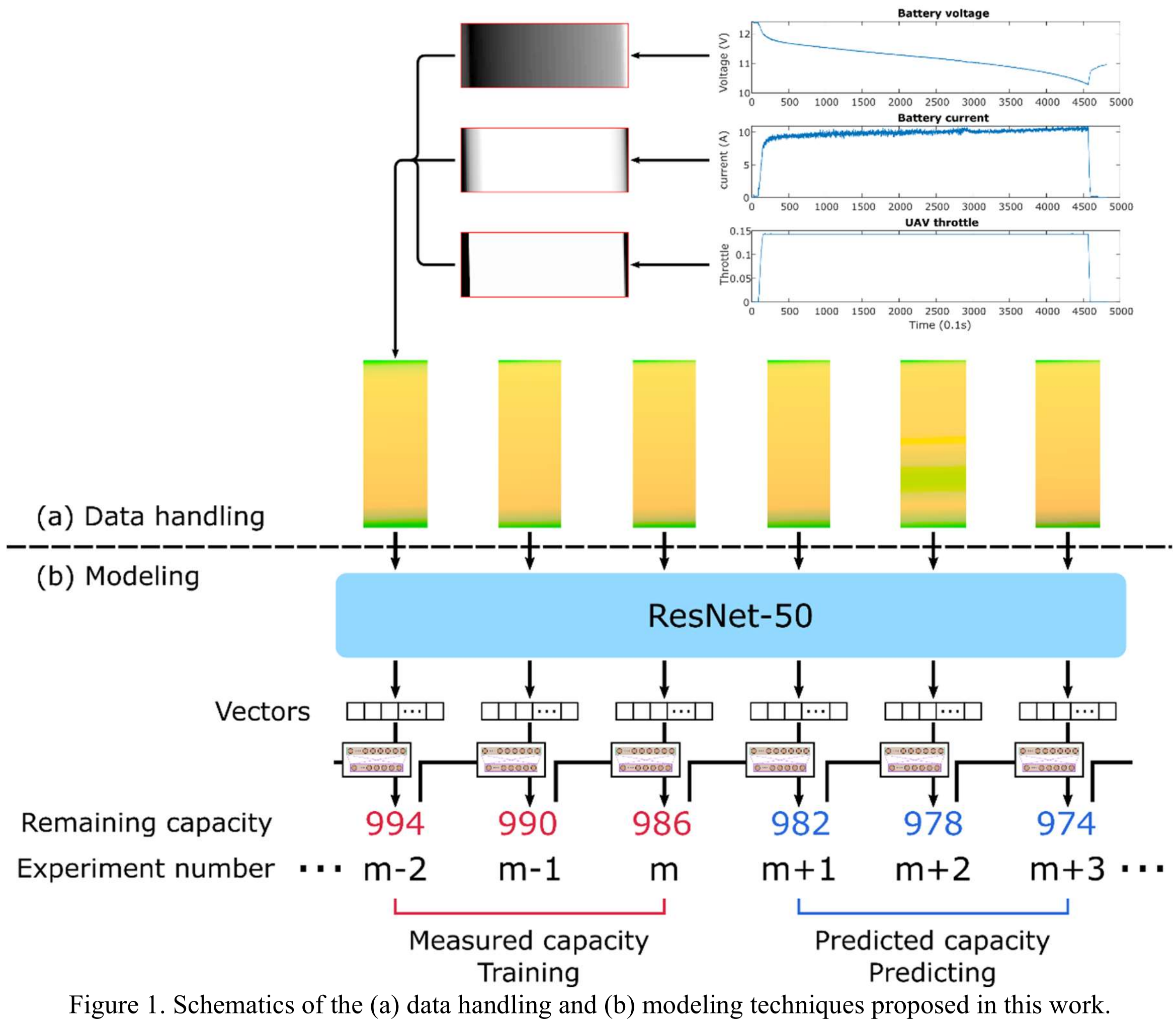

Figure 1. Schematics of the (a) data handling and (b) modeling techniques proposed in this work.

TABLE II. DESCRIPTIONS OF THE LIPO BATTERIES USED IN THE EXPERIMENTS

| Battery | Brand | Capacity | Cell size | Voltage | Discharge rate | Number of test batteries | Number of experiments |
|---|---|---|---|---|---|---|---|
| Type #1 | SIGP | 1100 mAh | 3S | 11.1 V | 20 C | 5 | 289 |
| Type #2 | RoaringTop | 2200 mAh | 3S | 11.1V | 25 C | 3 | 342 |

### C. *Machine Llearning*

ResNet-50 is a residual neural network with 50 layers that has demonstrated exceptional performance in image classification tasks. Its residual connections facilitate the training of deeper networks by mitigating the risk of gradient vanishing [16]. Pretrained using ImageNet, a large dataset of over 14 million images, ResNet-50 serves as a feature extractor to capture hierarchical representations from input images. Using a pretrained ResNet-50, a vector that has a size of 2048 was extracted from the image of each experiment (Figure 1(b)). This vector is fed into trainable fully connected layers (FCLs) to predict the remaining capacity after the experiment. Autoregression was implemented to utilize the time-series nature of the dataset and induce historical SoH knowledge to the model. Together with the extracted vector of the current experiment, the remaining capacity of the previous experiment was also fed into the FCLs as an input feature. The training and prediction phases use the measured remaining capacity and predicted remaining capacity of the previous experiment, respectively

## III. RESULTS

This case study aims to train an accurate SoH prediction model for battery Type #1, which has a smaller dataset than Type #2. The training set consisted of 4 Type #1 batteries, which had 203 experiments. The test set is composed of 1 Type #1 battery with 86 experiments. Knowledge transfer can be conducted from battery Type #2 to Type #1 to improve the model performance using transfer learning. The 342 experiments from Type #2 were used to pretrain the FCLs, which was then finetuned using the training set of Type #1. To show the impact of transfer learning, the FCLs were trained with and without transfer learning, respectively. Other ML models frequently utilized to predict battery SoH were also trained for comparison, including CNN and transformer.

TABLE III. STRUCTURE OF THE FULLY CONNECTED LAYERS

| Fixed hyperparameter/setting | Value/method |
|---|---|
| Activation | ReLU |
| Optimizer | Adam |
| Loss function | Mean squared error |
| Input and output layer dimensions | 2048 and 1 |
| Number of epochs | 2000 |
| **Tuned hyperparameters** | **Range** |
| Number of hidden layers | $(x \in \mathbb{N} \mid 1 \leq x \leq 3)$ |
| Number of neurons at each hidden layer | $(x \in \mathbb{N} \mid 20 \leq x \leq 200)$ |
| Initial learning rate | $(x \in \mathbb{R} \mid 0.001 \leq x \leq 0.1)$ |

Table III exhibits the settings and hyperparameter optimization method of the FCLs in Figure 1(b). 500 sets of hyperparameters were randomly sampled from the ranges defined in Table III using Latin hypercube sampling. 10% of the training set was distributed to the validation set. The model with the smallest validation error was chosen as the optimal model. The optimal hyperparameter sets were obtained when the FCLs were trained with and without transfer learning, respectively. Without transfer learning, the optimal hyperparameters were 150 hidden neurons in the first layer, 97 hidden neurons in the second layer, and an initial learning rate of 0.0256. With transfer learning, the optimal hyperparameters were 181 hidden neurons in the first layer, 157 hidden neurons in the second layer, and an initial learning rate of 0.0194. The learning curves of the two optimal models after 500 epochs are shown in Figure 2. The first 1000 epochs of transfer learning belong to the pretraining phase, where the FCLs were pretrained using the battery Type #2 data. Thereafter, the FCLs were finetuned using battery Type #2 training data. Right after the change of the training data, the loss curve had a slight increase, then quickly decreased and surpassed the loss curve of no transfer learning. At the 2000$^{th}$ epoch, the training loss of transfer learning was reduced to 0.0023, which was 28% lower than no transfer learning. The optimal hyperparameter sets of CNN and transformer were found using similar hyperparameter optimization methods.

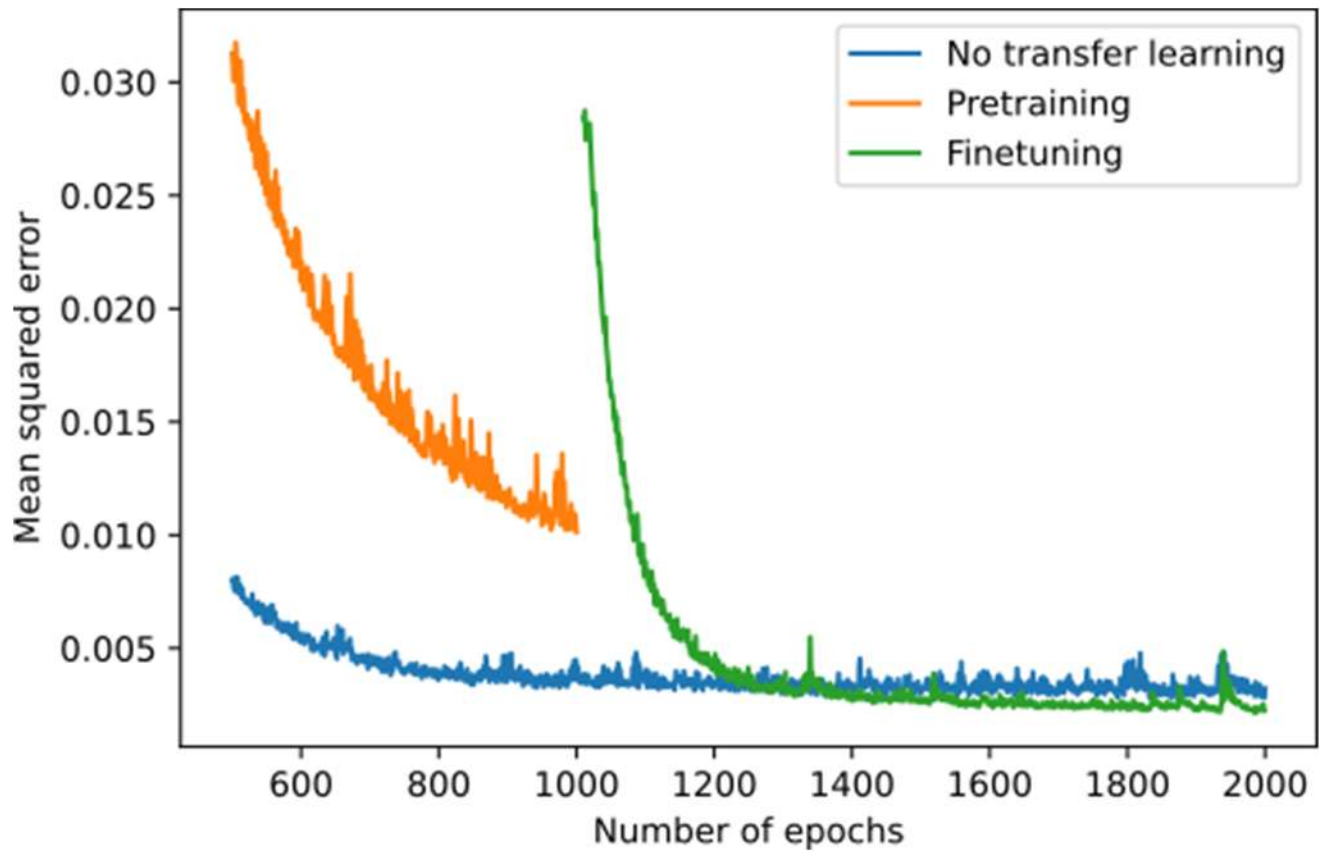

Figure 2. Loss curves of different training methods: no transfer learning and transfer learning (pretraining and finetuning).

The training and test performances were evaluated using mean squared error (MSE) and mean absolute percentage error (MAPE). The performances of the optimal ML models are shown in Table IV. The CNN model had the highest training MSE of 0.0072 because it had the smallest learning capacity among the candidate ML models. Thus, the test errors of the CNN model were higher than ResNet-50. Although transformers are state-of-the-art ML models, the optimal transformer model did not provide the best predictive performance. Besides, it was observed that the transformer model had significant overfitting to the training set. ResNet-50 with transfer learning had the lowest training and test errors among all the candidate ML models.

It can be observed that the pretrained ResNet-50 can extract salient representations of the input images converted from time-series experimental data. Subsequently, the ResNet-50 models can provide better predictive performance with scarce datasets. Additionally, transfer learning further improved the predictive performance by conducting knowledge transfer from similar batteries. Therefore, the strategies proposed in this research are effective when dealing with data scarcity in battery SoH prediction.

TABLE IV. TRAINING AND TEST PERFORMANCES OF THE OPTIMAL ML MODELS

| | **Training** | | **Test** | |
|---|---|---|---|---|
| **ML models** | ***MSE*** | ***MAPE*** | ***MSE*** | ***MAPE*** |
| **CNN** | 0.0072 | 0.0495 | 0.01019 | 0.0783 |
| **Transformer** | 0.0054 | 0.0325 | 0.01745 | 0.0922 |
| **ResNet-50 without transfer learning** | 0.0032 | 0.0258 | 0.0053 | 0.0347 |
| **ResNet-50 with transfer learning** | 0.0023 | 0.0176 | 0.0027 | 0.0226 |

## IV. CONCLUSIONS

This research proposed an ML-based battery SoH method for UAV PdM. The experiments were designed and conducted to collect the voltage, current, and throttle data from the battery discharging cycles using a UAV. To address data scarcity, the experimental data were converted to images so that a pretrained ResNet-50 could be utilized to extract representative features from the input. This research also conducted knowledge transfer to further improve the predictive accuracy of the model. As a result, an SoH prediction model with a test MAPE of 2.26% was obtained using a small dataset.

## ACKNOWLEDGMENT

This work is part of the PHUMS project established by VOZWIN Inc. Guidance and review provided by Fabio Bandera from VOZWIN Inc. has been a great help in the UAV design and experiments. We gratefully acknowledge VOZWIN Inc. for their ownership of the technology presented in this research.

## REFERENCES

[1] Stroe, A-I, Knap, Vaclav, and Stroe, D-I. "Comparison of lithium-ion battery performance at beginning-of-life and end-of-life." *Microelectronics reliability* Vol. 88 (2018): pp. 1251-1255.

[2] Wu, Yitao, Xue, Qiao, Shen, Jiangwei, Lei, Zhenzhen, Chen, Zheng, and Liu, Yonggang. "State of health estimation for lithium-ion batteries based on healthy features and long short-term memory." *IEEE Access* Vol. 8 (2020): pp. 28533-28547.

[3] Li, Weihan, Sengupta, Neil, Dechent, Philipp, Howey, David, Annaswamy, Anuradha, and Sauer, Dirk Uwe. "Online capacity estimation of lithium-ion batteries with deep long short-term memory networks." *Journal of power sources* Vol. 482 (2021): pp. 228863.

[4] Chen, James C, Chen, Tzu-Li, Liu, Wei-Jun, Cheng, CC, and Li, Meng-Gung. "Combining empirical mode decomposition and deep recurrent neural networks for predictive maintenance of lithium-ion battery." *Advanced Engineering Informatics* Vol. 50 (2021): pp. 101405.

[5] Rauf, Huzaifa, Khalid, Muhammad, and Arshad, Naveed. "Machine learning in state of health and remaining useful life estimation: Theoretical and technological development in battery degradation modelling." *Renewable and Sustainable Energy Reviews* Vol. 156 (2022): pp. 111903.

[6] Zahra, Nabila, Buldan, Rasis Syauqi, Nazaruddin, Yul Y, and Widyotriatmo, Augie. "Predictive maintenance with neural network approach for UAV propulsion systems monitoring." *2021 American Control Conference (ACC)*. pp. 2631-2636. 2021.

[7] Lee, Gyumin, Kwon, Daeil, and Lee, Changyong. "A convolutional neural network model for SOH estimation of Li-ion batteries with physical interpretability." *Mechanical Systems and Signal Processing* Vol. 188 (2023): pp. 110004.

[8] Shen, Sheng, Sadoughi, Mohammadkazem, Chen, Xiangyi, Hong, Mingyi, and Hu, Chao. "A deep learning method for online capacity estimation of lithium-ion batteries." *Journal of Energy Storage* Vol. 25 (2019): pp. 100817.

[9] Yang, Yixin. "A machine-learning prediction method of lithium-ion battery life based on charge process for different applications." *Applied Energy* Vol. 292 (2021): pp. 116897.

[10] Cheng, Gong, Wang, Xinzhi, and He, Yurong. "Remaining useful life and state of health prediction for lithium batteries based on empirical mode decomposition and a long and short memory neural network." *Energy* Vol. 232 (2021): pp. 121022.

[11] Park, Kyungnam, Choi, Yohwan, Choi, Won Jae, Ryu, Hee-Yeon, and Kim, Hongseok. "LSTM-based battery remaining useful life prediction with multi-channel charging profiles." *Ieee Access* Vol. 8 (2020): pp. 20786-20798.

[12] Fan, Yaxiang, Xiao, Fei, Li, Chaoran, Yang, Guorun, and Tang, Xin. "A novel deep learning framework for state of health estimation of lithium-ion battery." *Journal of Energy Storage* Vol. 32 (2020): pp. 101741.

[13] Kong, Depeng, Wang, Shuhui, and Ping, Ping. "State‐of‐health estimation and remaining useful life for lithium‐ion battery based on deep learning with Bayesian hyperparameter optimization." *International Journal of Energy Research* Vol. 46 No. 5 (2022): pp. 6081-6098.

[14] Li, Penghua, Zhang, Zijian, Grosu, Radu, Deng, Zhongwei, Hou, Jie, Rong, Yujun, and Wu, Rui. "An end-to-end neural network framework for state-of-health estimation and remaining useful life prediction of electric vehicle lithium batteries." *Renewable and Sustainable Energy Reviews* Vol. 156 (2022): pp. 111843.

[15] Gu, Xinyu, See, KW, Li, Penghua, Shan, Kangheng, Wang, Yunpeng, Zhao, Liang, Lim, Kai Chin, and Zhang, Neng. "A novel state-of-health estimation for the lithium-ion battery using a convolutional neural network and transformer model." *Energy* Vol. 262 (2023): pp. 125501.

[16] He, Kaiming, Zhang, Xiangyu, Ren, Shaoqing, and Sun, Jian. "Deep residual learning for image recognition." *Proceedings of the IEEE conference on computer vision and pattern recognition*. pp. 770-778. 2016.